# BCS-type mean-field theory for the $t$-$J$ model in the SU(2|1) superalgebra representation

Evgueny Kochetov
*Bogolyubov Theoretical Laboratory, Joint Institute for Nuclear Research, 141980 Dubna, Russia*

Marcin Mierzejewski
*Institute of Physics, Silesian University, 40-007 Katowice, Poland*
(Received 1 June 1999)

A simple version of the Bardeen-Cooper-Schrieffer (BCS)-type mean-field theory for the $t$-$J$ model is developed. The present approach *rigorously* treats the constraint of no doubly occupied states and invokes *two* local order parameters to implement spontaneous breaking of the global $U(1) \times U(1)$ symmetry. This is achieved by identifying the Hubbard operators with generators of the SU(2|1) superalgebra in the fundamental representation and employing the ensuing $CP^{1|1}$ parametrization for dynamical variables. As a result, an improved phase diagram $T_c(\delta)$ in the paramagnetic phase is obtained.

In 1987, P. W. Anderson made a suggestion[1] that the insulating state of a number of lanthanum copper oxides relevant for high-$T_c$ superconductivity is the resonating valence bond (RVB) state, a hypothetical spin liquid state proposed in 1973.[2] Such a state appears as a resonant disordered mixture of electron singlet pairs corresponding, to some extent, to the generalization of a disordered ground state of the one-dimensional (1D) quantum antiferromagnet to higher dimensions. Away from half filling the pre-existing spin singlet pairs can Bose condense giving rise to a charged superfluid state. To put this idea on quantitative grounds, Baskaran, Zou, and Anderson (BZA) (Ref. 3) worked out a variant of the BCS-type mean-field theory for the $t$-$J$ model, with the order parameter being the pair amplitude of a singlet along a nearest-neighbor bond. The authors seemed to properly recover a part of the phase diagram $T_c(\delta)$ for doping concentration $\delta > 0.05$ (for reasonable values of relevant parameters). The similar phase diagram has independently been derived by Ruckenstein, Hirschfeld, and Appel in their mean-field treatment of the extended $t$-$J$ model based on the slave-boson representation for the Hubbard operators.[4]

Although Anderson's notion of a spin liquid state seems quite reasonable, the RVB mean-field theory results in a number of unsatisfactory points, which may lead to some doubts when concerning its relevance for the high-$T_c$ phenomenon. First, $T_c$ starts out with a finite value $\sim J$, whereas a reasonable phase diagram must imply that $T_c(\delta=0)=0$. Second, away from half filling, $T_c$ increases to a maximum value which is, however, a few times higher than an experimental value. Third, the mean-field RVB theory predicts that $T_c(\delta)$ rapidly decreases beyond maximum until superconductivity is presumably destroyed, which would be in accord with a true phase diagram. On the one hand, actual computations reveal that the falloff of $T_c$ changes again at $\delta \sim 0.3$ for an increase, which certainly contradicts experimental results.

The purpose of this article is to demonstrate explicitly that the BZA mean-field theory does result in a reasonable phase diagram for the $t$-$J$ model, free of the above-mentioned inconsistencies, provided: (a) the constraint of no doubly occupied states is rigorously satisfied and: (b) order parameters to implement the $U(1) \times U(1)$ symmetry breaking are properly identified.

We start by expressing the $t$-$J$ model in terms of the Hubbard operators $X_i^{\sigma 0}$, defined as[5]

$$X_i^{\sigma 0} = c_{i\sigma}^\dagger (1 - n_{i,-\sigma}), \quad n_{i\sigma} n_{i,-\sigma} = 0,$$

where $c_{i\sigma}$ is the annihilation operator of an electron at site $i$ with spin $\sigma = \pm$, and $n_{i\sigma} \equiv c_{i\sigma}^\dagger c_{i\sigma}$. Then, the $t$-$J$ Hamiltonian takes on the form

$$H_{t-J} = -t \sum_{\langle i,j,\sigma \rangle} X_i^{\sigma 0} X_j^{0\sigma} + \text{H.c.} + J \sum_{\langle i,j \rangle} \left( \vec{Q}_i \vec{Q}_j - \frac{1}{4} n_i n_j \right), \quad (1)$$

where $\langle i,j \rangle$ denotes a summation over nearest-neighbor *non-repeated* bonds. The electron-spin operator

$$\vec{Q}_i = \frac{1}{2} \sum_{\sigma \sigma'} X_i^{\sigma 0} \vec{\tau}_{\sigma \sigma'} X_i^{0\sigma'}$$

enters the magnetic part of $H$ and $\vec{\tau} = (\tau^1, \tau^2, \tau^3)$ stand for the Pauli matrices.

Restriction of no double occupancy turns into identity $X^{00} + \Sigma_\sigma X^{\sigma\sigma} = I$. Note that $X^{\sigma 0}$ appears as a fermionic operator, whereas $X^{\sigma\sigma'}$ correspond to bosonic degrees of freedom. As a matter of fact, $X$ operators are closed into the SU(2|1) superalgebra in the lowest 3D (fundamental) representation (necessary details can be found in Ref. 6).

The BCS-type transition for the $t$-$J$ model in a superconducting state may occur as a result of spontaneous breaking of the $U_{N_e}(1) \times U_{Q_3}(1)$ global symmetry which corresponds to the conservation of the total electron number $N_e = \Sigma_i (X_i^{++} + X_i^{--}) =: \Sigma_{i,\sigma} n_{i\sigma}$ and the total spin projection operator $Q_3 = \frac{1}{2} \Sigma_i (X_i^{++} - X_i^{--})$. In order to imlpement spontaneous breaking of the $U_{N_e}(1)$ symmetry, consider the valence bond ''singlet'' pair creation and annihilation operator with the restriction of no double occupancy, $\sqrt{2} b_{ij}^\dagger = X_i^{+0} X_j^{-0} - X_i^{-0} X_j^{+0}$. Note that $b_{ii}^\dagger = 0$ (no doubly occupied states are possible within our formulation), whereas in the BZA theory one gets $b_{ii}^\dagger = c_{i\uparrow}^+ c_{i\downarrow}^+ \neq 0$. It is the important difference with respect to the BZA approach. A nonzero value of the order parameter $\sqrt{2}\langle b_{ij}\rangle =: \Delta_{ij} = \Delta$, implies that the $U_{N_e}(1)$ symmetry is spontaneously broken. Physically, it corresponds to the onset of the electron spin-singlet formation, with link-independent pair amplitude leading to an $s$-wave-like order parameter.

In the SU(2|1) path-integral representation, the $t$-$J$ partition function takes the form[6]

$$Z_{t-J} = \int_{CP^1} D\mu_{SU(2)} \int D\mu_F \exp[\mathcal{A}_{t-J}],$$

$$\mathcal{A}_{t-J} = \frac{1}{2}\sum_j \int_0^\beta dt \left( \frac{\dot{\bar{z}}_j z_j - \bar{z}_j \dot{z}_j}{1+|z_j|^2} + \dot{\bar{\xi}}_j \xi_j - \bar{\xi}_j \dot{\xi}_j \right) - \int_0^\beta H_{t-J}^{cl} dt. \quad (2)$$

The classical Hamiltonian which enters the action reads

$$H_{t-J}^{cl} = -t\sum_{\langle i,j\rangle} \frac{\xi_i \bar{\xi}_j (1+\bar{z}_i z_j)}{\sqrt{(1+|z_i|^2)(1+|z_j|^2)}} + \text{H.c.}$$

$$-\mu\sum_i (1-\bar{\xi}_i\xi_i) - \frac{J\Delta}{2}\sum_{\langle i,j\rangle} \frac{\xi_i \xi_j (\bar{z}_j - \bar{z}_i)}{\sqrt{(1+|z_i|^2)(1+|z_j|^2)}}$$

$$+ \text{H.c.} + J\Delta^2 NZ/4, \quad (3)$$

where we have explicitly introduced a chemical-potential term $-\mu N_e$ to control the occupation number. Even and odd Grassmann variables $z_i$ and $\xi_i$ parametrize superspace SU(2|1)/U(1|1) = $CP^{(1|1)}$, the $N=1$ supersymmetric extension of complex projective line $CP^1$. From Eq. (2) it follows that $\xi_i$ correspond to the spinless fermion degrees of freedom, while $z_i$ represent pure SU(2) spins.[6] Operators $X_i^{\sigma 0}$, which describe real electron excitations (provided there are no doubly occupied sites) are thus replaced by auxiliary dynamical fields $z_i$ and $\xi_i$ which, to some extent, keep track of the electron spin and charge degrees of freedom, respectively. Relevant elementary excitations are usually termed spinons and holons. The $CP^{1|1}$ language provides an adequate description of the spinon-holon excitations inherent to a doped quantum spin liquid state,[7] though of course not unique. For instance, the SU(2) formulation of the $t$-$J$ model to discribe the underdoped regime has recently been successfully developed.[8]

Equation (3) is invariant under global $U_{Q_3}(1)$ transformations $\xi_i \to e^{i\theta}\xi_i$, $z_i \to z_i e^{2i\theta}$. To properly identify an order parameter associated with the $U_{Q_3}(1)$ symmetry breaking, a certain fixed value of $z$ field, $z_i^{(0)}$, might be ascribed to any lattice site $i$, which leads to the spontaneous breaking of the $U_{Q_3}(1)$ symmetry by the local order parameter

$$\Lambda_{ij} = -\Lambda_{ji} = \frac{(z_i^{(0)} - z_j^{(0)})}{\sqrt{(1+|z_i^{(0)}|^2)(1+|z_j^{(0)}|^2)}}.$$

The latter can be viewed as a spinon-singlet amplitude. To see this, one may note that

$$\Lambda_{ij} = \Psi_{s_3=-1/2}(z_i)\Psi_{s_3=1/2}(z_j) - \Psi_{s_3=-1/2}(z_j)\Psi_{s_3=1/2}(z_i),$$

where $\Psi_{s_3=\pm 1/2}(z)$ is a spinon wave function in the $z$ representation:

$$\Psi_{s_3=\pm 1/2}(z) := \langle s_3 = \pm 1/2|z\rangle, \quad \hat{S}_{(3)}|s_3\rangle = s_3|s_3\rangle,$$

$$|z\rangle = (1+|z|^2)^{-1/2}\exp\{zQ_-\}|s_3=1/2\rangle.$$

To proceed, we suggest the simplest mean-field treatment of the RVB state based on the SU(2|1) path-integral representation (2). Essentially, as the spinon field $z_i$ can fluctuate between two values $\pm\alpha \in C$, we assign a classical fluctuating field, $z_i^{(0)} = \pm\alpha$, to every lattice site. If this is the case, then there is no sense in introducing any type of the antiferromagnetic sublattices. Of course, a certain (highly simplified) scheme for these fluctuations has to be imposed by hand, which may be interpreted as a snapshot in the path integral. Explicitly, we put $z_i^{(0)} - z_{j(i)}^{(0)} = \alpha[1-(-1)^{i\vec{\pi}(\vec{R}_i-\vec{R}_{j(i)})}] = 2\alpha$, where index $i$ runs over all the lattice sites, whereas $j(i)$ is defined by

$$\vec{R}_{j(i)} = \vec{R}_i + \vec{n}, \quad \vec{n} > 0, \quad (4)$$

with $\vec{n}$ being the unit vector in the lattice axes directions. To implement this ansatz into path integral (2), $D+1$ fields $z_j, z_{j(i)}$ at every site $j$ are considered instead of one field $z_j =: z_{j(j)}$, where $D$ indices $i$ are defined at a given $j$ by Eq. (4). Integration is then understood over all the auxiliary fields, with $D$ constraints $z_j + z_{j(i)} = 0$ being imposed.

Once this is assumed, the quantum counterpart of Eq. (3) becomes (in the momentum representation)

$$H_{t-J} = -\sum_{\mathbf{k}} t_{\mathbf{k}}(\alpha) f_{\mathbf{k}}^+ f_{\mathbf{k}}$$

$$-\mu\sum_{\mathbf{k}} (1-f_{\mathbf{k}}^+ f_{\mathbf{k}}) + iS^+(\alpha)J\Delta$$

$$\times \sum_{\mathbf{k}} f_{\mathbf{k}} f_{-\mathbf{k}} \beta_{\mathbf{k}} - iS^-(\alpha)J\Delta$$

$$\times \sum_{\mathbf{k}} f_{-\mathbf{k}}^+ f_{\mathbf{k}}^+ \beta_{\mathbf{k}} + J\Delta^2 NZ/4, \quad (5)$$

where $f^{cl} := \xi$ and $t_{\mathbf{k}}(\alpha) = 2tS^3(\alpha)\gamma_{\mathbf{k}}$.

Hamiltonian (5) is of BSC type, except that spinless fermions are involved rather than the true fermionic operators. This is an obvious consequence of the no double occupancy restriction. To proceed, a paramagnetic phase is to be considered, $\langle Q_i^3\rangle = 0$, where the thermodynamical averaging is understood with respect to the Hamiltonian (5). The above equation relates the hole concentration $\delta = (1/N)\sum_i \langle f_i^+ f_i\rangle$ to the spinon variable $\alpha$ in the following way: $\delta = (1-|\alpha|^2)/(1+|\alpha|^2) = -2S_3$ (it is sufficient to consider $|\alpha| \in [0,1]$, otherwise one can make a change $\alpha \to 1/\alpha$ and $t \to -t$). As a consequence, no true long-range magnetic order emerges. We have considered anisotropic order parameter: $\Delta(\mathbf{q}) = \Delta(\cos q_x \pm \cos q_y)$ with $+(-)$ corresponding to extended $s$-wave ($d$-wave) symmetry. One obtains the following system of equations which determine the order parameter and the chemical potential:

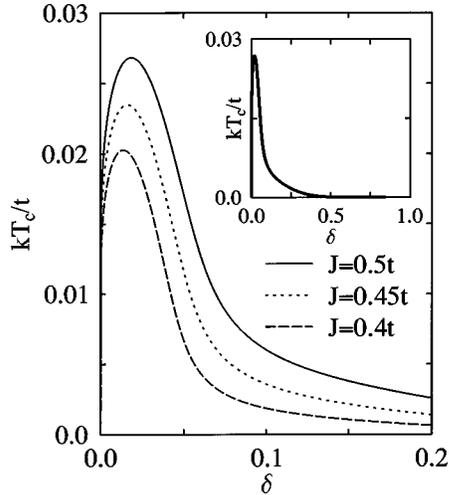

FIG. 1. The critical temperature as a function of concentration of holes. The inset shows $T_c(\delta)$ calculated for $J=0.5t$.

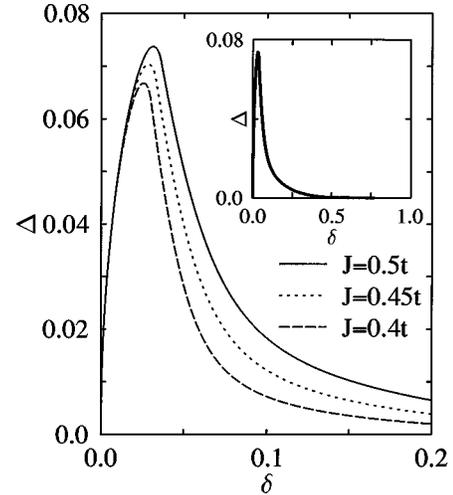

FIG. 2. Magnitude of the order parameter at $T=0$ as a function of concentration of holes. The inset shows $\Delta(\delta)$ at $T=0$ calculated for $J=0.5t$.

$$\frac{1}{N}\sum_{\mathbf{k}}\frac{\tanh(E_{\mathbf{k}}\beta/2)}{E_{\mathbf{k}}}(\beta_{\mathbf{k}}^{(s,d)})^2=\frac{Z}{J}, \quad (6)$$

$$\frac{1}{2N}\sum_{\mathbf{k}}\frac{\tanh(E_{\mathbf{k}}\beta/2)}{E_{\mathbf{k}}}[t_{\mathbf{k}}(\delta)-\mu]=\delta-1/2, \quad (7)$$

where

$$E_{\mathbf{k}}^2=J^2\Delta^2(\beta_{\mathbf{k}}^{(s,d)})^2(1-\delta^2)+[t_{\mathbf{k}}(\delta)-\mu]^2, \quad (8)$$

$t_{\mathbf{k}}(\delta)=-2t\delta\gamma_{\mathbf{k}}$, $\gamma_{\mathbf{k}}=\cos k_x+\cos k_y$, $\beta_{\mathbf{k}}^s=\sin k_x+\sin k_y$ for extended $s$-wave and $\beta_{\mathbf{k}}^d=\sin k_x-\sin k_y$ for $d$-wave symmetry of the order parameter.

Note that setting in the above equations $\Delta=0$ does not correspond to a normal phase which by definition is the state with the unbroken $U_{N_e}(1)\times U_{Q_3}(1)$ symmetry. As a consequence, Eqs. (6)–(8) result at $\Delta=0$ in a nonmonotonic behavior of chemical potential $\mu$ as a function of $\delta$, which is in a sense a generic feature of a state with broken symmetry. Within our approach the normal phase implies that both equations $\Delta=0$ and $\Lambda=0$ hold, which yields a monotonic though trivial dependence $\mu(\delta)$.

In contradistinction to the standard RVB formulation, in our approach the exclusion of double occupancy of lattice sites is explicitly taken into account. This feature shows up in the equation for the chemical potential (7) and results in a different occupation number dependence of pairing correlations. In order to illustrate this difference one can consider Eqs. (6) and (7) in the limit $\delta\to 0$ and $\Delta\to 0$, which allows for evaluation of the critical temperature for vanishing concentration of holes. In the BZA approach the case $\delta\to 0$ corresponds to the half filled band. One obtains $\mu=0$ and $kT_c=J/(ZN)\Sigma_{\mathbf{k}}\gamma_{\mathbf{k}}^2\neq 0$. However, for spinless fermions $\delta\to 0$ is related to entirely filled band. It can be easily proved that for vanishing concentration of holes Eqs. (6) and (7) lead to $\mu=J/(ZN)\Sigma_{\mathbf{k}}(\beta_{\mathbf{k}}^{(s,d)})^2$ and $T_c=0$ for both (extended $s$-wave and $d$-wave) symmetries. Therefore one may expect that the critical temperature achieves the maximum for finite doping. It has been confirmed by numerical calculations. Figures 1 and 2 show the critical temperature and the magnitude of the order parameter at $T=0$ as a function of concentration of holes. Both $T_c$ and $\Delta(T=0)$ do not depend on the symmetry of the RVB state. In our approach the RVB pairing mechanism is effective only for small concentration of holes $\delta < 0.3$—0.4 and leads to much lower (and therefore more realistic) values of the critical temperature than in the original BZA formulation. In the entire region of concentration of holes $T_c(\delta)$ reproduces, at least qualitatively, the experimental data for high-temperature superconductors. The similar behavior for the $d$-wave superconducting order parameter $\Delta(T=0)$ as a function of $\delta$ has been obtained in Ref. 9, within a different framework, based on a variational method for a projected BCS trial wave function in the Gutzwiller approximation.

It is worth mentioning that a few earlier attempts at developing mean-field theories directly in terms of the $X$ operators have been made. Since the slave particle representations for the Hubbard operators necessarily imply that the constraint of no double occupancy is to be taken into account (see, e.g., Ref. 10, and references therein), and in view of the fact that its mean-field treatment leads to an uncontrollable error,[12] some attempts to derive specific *exact* spin-fermion representations for Hubbard operators free of any constraints have been carried out.[11] Nevertheless, those representations, exact at $\delta=0$, have been proven false away from half filling.[6]

On the other hand, the mean-field theory based on the so-called projection technique for Green functions in terms of Hubbard operators has been developed.[13] This approach heavily relies on the so-called Hubbard-I approximation, $\langle X_i^{\sigma\sigma}\rangle=n/2$, which seems to impose a rather strong restriction. As a matter of fact, the mean-field theory based on both the projection technique and Hubbard-I approximation proves, in general, inconsistent with the underlying SU(2|1) superalgebra identities, which casts certain doubts on the results obtained within that approach. In particular, it has been conjectured[13] that the $d$-wave superconducting pairing in the $t$-$J$ model follows just from the SU(2|1) algebra identities, whereas our consideration, also based on the $X$-operator algebra, does not distinguish between the $d$- and $s$-wave pairings. This apparent distinction deserves a comment.

The authors of Refs. 13 infer their conclusion from the equation

$$\langle X_i^{\sigma 0} X_i^{\bar{\sigma} 0} \rangle = 0, \quad \bar{\sigma} = -\sigma \tag{9}$$

which, in their opinion, enforces the requirement of no doubly occupied states at a site. In fact, the operator identity

$$X_i^{\sigma 0} X_i^{\bar{\sigma} 0} = 0 \tag{10}$$

that follows from the definition of the Hubbard operators and does eliminate two-particle states, is not equivalent to Eq. (9), but rather implies that Eq. (9) must hold *irrespective* of a particular way of averaging (in our treatment, for instance, $\langle X_i^{+0} X_i^{-0} \rangle \sim \langle \xi_i^2 \rangle \equiv 0$). In other words, Eq. (9), as it appears in Ref. 13, must be obeyed for *any* function $\Delta_\mathbf{q}$, and hence can show no preference to a particular choice of the superconducting order parameter. If, on the other hand, Eq. (9) holds only for a certain specific set of functions $\Delta_\mathbf{q}$—which is just the case in Ref. 13—Eq. (10) cannot be obeyed and hence no properly defined Hubbard operators are actually involved. From a physical standpoint, this results in an uncontrollable error and consequently in unreliable conclusions.

To summarize, we suggest a more refined treatment of the BZA mean-field theory for the t-J model based on a rigorous imposition of the no double occupancy constraint and a proper identification of the local order parameters relevant for BCS-type transition. As a result, a reasonable phase diagram in the paramagnetic phase is obtained.

The authors are indebted to Janusz Zielinski for helpful discussions. E.K. also thanks the Institute of Physics at the University of Katowice for hospitality extended to him during a final stage of this work.

[1] P.W. Anderson, Science **235**, 1196 (1987).
[2] P.W. Anderson, Mater. Res. Bull. **8**, 153 (1973).
[3] G. Baskaran, Z. Zou, and P.W. Anderson, Solid State Commun. **63**, 973 (1987).
[4] A.E. Ruckenstein, P.J. Hirschfeld, and J. Appel, Phys. Rev. B **36**, 857 (1987).
[5] J. Hubbard, Proc. R. Soc. London, Ser. A **285**, 542 (1965).
[6] E.A. Kochetov and V.S. Yarunin, Phys. Rev. B **56**, 2703 (1997).
[7] R.B. Laughlin, Science **242**, 525 (1988).
[8] P.A. Lee, N. Nagaosa, T.K. Ng, and X.G. Wen, Phys. Rev. B **57**, 6003 (1998).
[9] F.C. Zhang, C. Gros, T.M. Rice, and H. Shiba, Supercond. Sci. Technol. **1**, 36 (1988).
[10] P.A. Lee and N. Nagaosa, Phys. Rev. B **46**, 5621 (1992).
[11] L. Richard and V.Yu. Yushankhai, Phys. Rev. B **47**, 1103 (1993); Y.R. Wang and M.J. Rice, *ibid.* **49**, 4360 (1994).
[12] S. Feng, Z.B. Su, and L. Yu, Phys. Rev. B **49**, 2368 (1994).
[13] N.M. Plakida, V.Yu. Yushankhai, and I.V. Stasyuk, Physica C **160**, 80 (1989); V.Yu. Yushankhai, N.M. Plakida, and P. Kalinay, *ibid.* **174**, 401 (1991); N.M. Plakida, Philos. Mag. B **76**, 771 (1997); N.M. Plakida and V.S. Oudovenko, Phys. Rev. B **59**, 11 949 (1999).